\documentclass[twocolumn,showpacs,preprintnumbers,amsmath,amssymb,letterpaper]{revtex4}
\usepackage{graphicx}
\usepackage{dcolumn}
\usepackage{bm}
\usepackage{amssymb}
\usepackage{epsfig}

\newcommand{\be}{\begin{equation}}
\newcommand{\ee}{\end{equation}}
\newcommand\beq{\begin{eqnarray}}
\newcommand\eeq{\end{eqnarray}}
\newcommand\eqn[1]{\label{eq:#1}}
\newcommand\eq[1]{Eq. (\ref{eq:#1})}

\newcommand{\vev}[1]{\langle #1 \rangle}

\newcommand{\eV}{{\rm ~eV }}

\newcommand{\CN}{{\cal N}}

\newcommand{\CR}{{\cal R}}

\newcommand{\CZ}{{\cal Z}}
\newcommand{\CL}{{\cal L}}

\newcommand{\Tr}{{\rm Tr\,}}

\newcommand{\mybar}[1]%
        {\kern 0.8pt\overline{\kern -0.8pt#1\kern -0.8pt}\kern 0.8pt}
\newcommand{\sla}[1]%
        {\raise.15ex\hbox{$/$}\kern-.57em #1}
\begin{document}

\preprint{INT-PUB-11-061}

\title{Spacetime as a topological insulator: Mechanism for the origin of the
    fermion generations}

\author{David B. Kaplan}
\email{dbkaplan@uw.edu}

\author{Sichun Sun}
 \email{sichun@uw.edu}

 \affiliation{Institute for Nuclear Theory, Box 351550, Seattle, WA 98195-1550, USA}

 \date{\today}

 \begin{abstract}
We suggest a mechanism whereby the three generations of quarks and leptons correspond to surface modes in a five-dimensional theory. These modes arise from a nonlinear fermion dispersion relation in the extra dimension, much in the same manner as fermion surface modes in a topological insulator or lattice implementation of domain wall fermions. We also show that the topological properties can persist in a deconstructed version of the model in four dimensions.

\end{abstract}

\pacs{12.15.Ff, 14.60.Pq, 73.20.-r, 11.10.Kk }
%
%
\maketitle

\section{Introduction}

It remains a mystery why there are three particle generations in the standard model and why they have the observed  pattern of masses and mixing angles, despite many attempts at explanation, experimental evidence for flavor physics beyond the standard model being limited to neutrino masses.  The smallness of neutrino masses and the absence of flavor-changing neutral currents and  electric dipole moments all suggest that the origin of flavor lies enigmatically at very short distance.    A well explored theoretical program is to  assume the existence of three generations and guess at   textures for the mass matrices  which can be justified by a hierarchy of flavor symmetry breaking --- an ambiguous exercise given  the lack of experimental flavor probes in the right-handed fermion sector.  After pioneering work involving abelian flavor symmetries \cite{Froggatt:1978nt},  numerous models were also introduced with non-abelian flavor symmetries possessing three dimensional irreducible multiplets to justify the existence of three generations of quarks and leptons.  While this general approach can boast of qualitative successes, no models have emerged that are particularly compelling.

Composite and extra dimension models  are a natural place to look for an explanation for flavor and the number of generations:  both generically contain towers of states, and one can arrange that only three generations are light.  Composite models must typically rely on gauge dynamics to explain the  origin of three generations, as in  Ref.~\cite{Kaplan:1997tu}, while extra dimension models often rely on the Dirac equation having three  zero modes in certain background fields of nontrivial topology Ref.~\cite{Aguilar:2006sz,Beasley:2008dc,Guo:2008ia}. In composite models, the Yukawa matrices  of the standard model are due to complex interactions between constituents, and at best their texture can be predicted; in extra dimension models, the Yukawa matrices can be computed from wave function overlap integrals in the transverse space (see, for example, \cite{Katz:1996xe,Grossman:1999ra,Gherghetta:2000qt,Aguilar:2006sz,Fitzpatrick:2007sa,Guo:2008ia,Perez:2008ee,Heckman:2008qa,Beasley:2008dc,Beasley:2008kw,Randall:2009dw,Guo:2009gb}). To the extent a gauge/gravity duality pertains, it is possible that these two very dissimilar descriptions could be related.

In this Letter we consider an interesting phenomenon observed with lattice domain wall fermions, where the number of massless fermions bound to the surface of a semi-infinite fifth dimension depended discontinuously on the fermion dispersion relation, and hence on coupling constants in the action  \cite{Kaplan:1992bt,Jansen:1992tw}.  It was subsequently shown that the number of light ``families" could be understood as a topological property of the five-dimensional (5D) fermion dispersion relation in momentum space 
\cite{Golterman:1992ub}. That is because the number of 4D massless surface modes is directly related to the quantized coefficient of the Chern-Simons operator obtained by integrating out the heavy bulk fermions, following the analysis in Ref.~\cite{Callan:1984sa}; and this coefficient is obtained from a one-loop Feynman diagram which computes a momentum-space winding number associated with the fermion propagator.
This phenomenon was first discussed in the   classification of fermion modes in liquid helium \cite{grinevich1988topology}  and is the same phenomenon that defines topological insulators \cite{kane2005quantum,Hasan:2010xy,2008PhRvB..78s5424Q}. We consider here that  the replication of quark and lepton families  we observe in the standard model arise in such a manner (see \cite{Volovik:2011kg} for related speculations);  an attractive feature of the mechanism is that while the number of light families is determined topologically, their  transverse wave functions in the extra dimension are all different and dynamically determined, allowing interesting mass mixing without overly restrictive family symmetries. As topology in momentum space depends on the large momentum behavior of the fermion dispersion relation,  such models are forced to confront UV physics, and cannot simply rely on an effective field theory description.  Therefore after describing the general mechanism and providing a phenomenological toy example, we look at various UV completions that can give rise to a well-defined low energy theory.

\section{Multiple zero modes}

We start by considering fermions in 5D,  with an inverse Euclidian propagator which respects 4D Lorentz invariance
\beq
iG^{-1}(p_\mu, p_5) = iZ_\mu(p) \gamma^\mu + i Z_5(p_5)\gamma^5 -\Sigma(p,p_5)\ ,
\eqn{disp}\eeq 
corresponding to a plane wave $u(p) \exp(i p_a x^a)$, where $p_a = \{p_\mu, p_5\}$ is the 5-momentum and $u$ is a Dirac spinor. We assume Hermitean gamma matrices with $Z_\mu$ and $Z_5$ being real, odd functions of momentum, and $\Sigma$  a real, even function, so that $G^{-1}$ corresponds to Hermitean derivative interactions in the fermion action. $G^{-1}$ will typically have one or more zeros for real $p^\mu$ and some complex value for $p_5$, and a $u$ spinor which is an eigenstate of $\gamma_5$.   For the generic case $\Im[p_5]\ne 0$, this pole in $G$ implies a wave function growing exponentially in one of the $x_5$ directions; because of the symmetry of $Z$ and $\Sigma$ under $p_5\to -p_5$, if there is chiral solution to $G^{-1}=0$ with one sign for $\Im[p_5]$, there is also a solution with the opposite chirality and the opposite sign for $\Im[p_5]$.  Such solutions are not normalizable; however, if translation invariance in $x_5$ is destroyed, either by a boundary or explicit localized  $x_5$ dependence in the action, it is possible that some chiral solutions are normalizable and retained in the Hilbert space --- confined to the boundary or defect --- while modes of the opposite chirality remain non-normalizable and are discarded.  The result is an effective 4D theory at low energy with chiral fermions; this is just a generalization of the domain wall fermion zero mode discovered by Jackiw and Rebbi \cite{Jackiw:1975fn}.  If there are multiple solutions to $G^{-1}=0$, then the 4D theory will have multiple families, distinguished by different transverse wave functions in the 
extra dimension. While these wave functions will depend continuously on parameters in the 5D Lagrangian, the number of families can only depend discontinuously on those parameters; this makes the number of families look like a topological number, as indeed it is \cite{Golterman:1992ub}. For example, for 5D lattice domain wall fermions one finds $\Sigma=[m + r\sum_{a=1}^5 (\cos p_a-1) ]$, $Z_a = \sin p_a$,   $r$ being a free parameter. It was shown in Ref.~\cite{Jansen:1992tw}  that when the extra dimension is made semi-infinite, chiral zero modes exist at the boundary, where the number of families jumps through the binomial coefficients $\{1,4,6,4,1\}$  as the ratio $(m/r)$ is tuned through multiples of two, alternating in chirality with each jump.
We suggest here that the three families observed in nature might arise in such a manner.

\subsection{A toy model}

For a toy model with three generations we start with the simple and unjustified assumption that the dispersion relation  \eq{disp} is given by
\beq
iG^{-1} = p_\mu \gamma^\mu + i p_5(1+c_1 p_5^2) \gamma_5 - m(1+ c_2 p_5^2)
\eqn{dispii}
\eeq
where $m$, $c_1$ and $c_2$ are real and chosen so that for $p_\mu=0$, $G^{-1}$  has three roots, all of a given chirality,  occurring at $p_5 = i\kappa$ with
$
 \kappa_1=a$, $\kappa_2=(b+ic)$, $\kappa_3=(b-ic)$, where $a,b,c$ are real, positive numbers. These three roots correspond to transverse wave functions for the zero modes of the form $\phi_i(y) = \exp(-\kappa_i y)$. For now we consider the extra dimension to be semi-infinite, and we ignore gravity.

Expanding  the 5D fields in this non-orthonormal set of transverse zero mode wave functions yields the low energy 4D theory with off-diagonal kinetic terms, $\CZ_{ij} \psi^\dagger_iD_\mu \sigma^\mu\psi_j$, where $\psi_i(x)$ is the chiral  spinor associated with the $i^{th}$ zero mode, $i,j=1,2,3$, and the wave function mixing matrix $\CZ$ is given by the  overlap of transverse wave functions,
  \beq
{\cal Z}_{ij} = \int_0^\infty  dy\,  \phi^*_i(y) \phi_j(y) = \frac{1}{\kappa_i^*+ \kappa_j}\ .
\eqn{zmat}\eeq
4D gauge invariance requires there to be a $y$-independent mode for the $W$ bosons, and so weak currents and ${\cal Z}$ will both be diagonal in the flavor basis.  This is more apparent if one discretizes the extra dimension and considers it as a flavor index, so $y$-independence of the $W$  is equivalent to the statement that the $W$ couples to each flavor in the UV theory with the same coupling constant.

The natural starting point for construction of a low energy description of the world  is to assume  one Dirac field in the 5D theory  for each  Weyl field of the standard model: $Q, U, D, L, E$ with $SU(3)\times SU(2)\times U(1)$ quantum numbers $(3,2)_{1/6}$, $(3,1)_{2/3}$, $(3,1)_{-1/3}$, $( 1,2)_{-1/2}$ and $(1,1)_{-1}$ respectively.  We assume that $Q$ and $L$ only has left-handed chiral zero modes $q_i$, $\ell_i$, while $U$, $D$ and $E$ have only right-handed zero modes, $u_i$, $d_i$, $e_i$.
These fields can then couple to the Higgs in the 5D theory as $\mybar U H Q$, etc. (Alternatively one could replace $U$, $D$, $E$ by conjugate fields $U^c$, $D^c$ and $E^c$ with Higgs couplings $U^c H C_5 Q$, etc, where $C_5$ is the 5D charge  conjugation matrix, and assume a 5D dispersion relation where all zero modes are left-handed.)

A problem with the above implementation is that, with a single Yukawa coupling for each 5D field, it is not possible to include weak CP violation.  We therefore make the Higgs sector the origin of $CP$ violation in the UV theory,
which requires introducing additional scalars.  We assume a two-doublet model with a relative $CP$-violating phase in their vacuum expectation values (vevs) arising from explicit $CP$-violation in the Higgs potential.  To avoid large flavor changing neutral currents, the theory is given a discrete symmetry (softly broken) to ensure that $H_u$ and $H_d$  couple solely to up-type and down-type quarks respectively  \cite{Glashow:1976nt}.   Furthermore we must assume that these are bulk fields and that Higgs vacuum expectation values are  $y$-dependent; this is the same model and mechanism commonly used to introduce $CP$ violating bubble walls in theories of electroweak baryogenesis \cite{Cohen:1993nk}.
 The Yukawa interactions of the quarks  in the 5D theory are given by
$(y_U \mybar U  H_u Q+ y_D \mybar D H_d Q + \text{h.c.})$,
 In principle there could be higher derivative Yukawa-like interactions, but we assume they are zero at the UV scale.

As with the fermion zero modes, we assume an exponential form for the 5D profile of the Higgs vevs. By means of a hypercharge gauge transformation we can make $H_d$ vev to be real and put all the $CP$ violating phase into the $H_u$:
$
\vev{H_d}=ve^{-y} \sin\beta$,
$\vev{H_u} =ve^{-(h_r + i h_i)y}\cos\beta$,
where  we have chosen the scale of the coordinate $y$ so that the exponent in the profile of $\vev{H_d}$ equals one, while the real parameters $h_r$ and $h_i$ characterize the profile of $\vev{H_u}$.
Integrating over the coordinate $y$  then gives rise to conventional 4D mass matrices, such as
$
[M_U]_{ab} \propto( y_U v\cos\beta/\sqrt{2})
(\kappa_{Q,a}+\kappa_{U,b}+h_r+i h_i)^{-1} $.

 In our model we choose Higgs couplings to the leptons of the form
 \beq
y_E \mybar L \widetilde H_u E+ \frac{1}{\Lambda}(L\widetilde H_d)^TC_5(L\widetilde H_d) + \text{h.c}
\eqn{maj}
\eeq
where $\widetilde H = \sigma_2 H^*$; $\Lambda$ has dimensions of mass and controls the size of the resulting Majorana neutrino masses.
 After integrating over the coordinate $y$ the Yukawa interactions give rise to mass matrices 
which are simple functions of the phenomenological $\kappa$ parameters; from these and the wave function matrices \eq{zmat} it is straightforward computation to determine all the masses and mixing angles.

The point of this toy model is not to present a full theory of 5D physics, but only want to see whether a model based on this topological insulator mechanism could agree with experimental data on flavor physics to high accuracy.
The model described has 21 real parameters: $a$, $b$ and $c$ for the $Q$, $U$,  $D$,  $L$ and $E$ fields, $h_r$ and $h_i$ for  $\vev{H_u}$, the three Yukawa couplings $y_U$, $y_D$, $y_E$, and the scale $\Lambda$ characterizing the neutrino coupling to the Higgs.    These are fit to 18 data:  the nine quark and charged lepton masses, three CKM angles and one phase, two neutrino $\Delta m^2$ values, and three neutrino mixing angles.  Ignoring uncertainties in the data, in general one would expect some number of disconnected three-dimensional manifolds in parameter space where the model agrees with all the data  ---  that number possibly being zero; lower dimensional spaces of solutions would generally require fine tuning.
In searching for  solutions, we ignore radiative corrections in our model (such as running of the Yukawa couplings and masses, or radiative generation of higher derivative operators), and fit the parameters to data currently available from  the Particle Data Group  \cite{nakamura2010review}, augmented with recent evidence for nonzero $\theta_{13}$ in the neutrino sector  \cite{Fogli:2011qn,An:2012eh}.

One might think that this model, with more parameters than data, could not be predictive.  However, when we numerically map out the manifold of solutions consistent with the data (including experimental and theoretical errors quoted in the PDG) we find (i) we always get a normal, nondegenerate hierarchy, with $m_3$ in the narrow range $0.048\eV \le m_3 \le 0.051\eV$; (ii) solutions do not favor maximal mixing for $\nu_2-\nu_3$; (iii) we find $J_\nu$ in the narrow range $-0.023 \le J_\nu\le -0.014$.  What has happened is just that the three-dimensional manifold of predictions form this constrained model maps onto a narrow range of physical properties, so that it is in fact somewhat predictive. Nevertheless, one would like to understand how to construct a model which is both more predictive and is well defined in the UV.

\section{UV completion: Little Flavor}

A general method for constructing a UV completion for extra dimension models is to discretize the extra dimensions while keeping the 4D world continuous (deconstruction) \cite{ArkaniHamed:2001ca,ArkaniHamed:2001nc}.  In such an approach, the extra coordinate for bulk fermions essentially becomes a flavor index. A 4D deconstructed version of a free theory with dispersion relation similar to \eq{dispii} is readily obtained by discretizing an infinite extra dimension with defect at site $n=0$.  In this case we have an infinite number of flavors of 4D fermions with the mass matrix
$\CL_m = (\bar\psi_L M\psi_R + \text{h.c.})$,
where $M$ is an infinite matrix representing the fifth dimension differential operator. For example:
\beq
(Mv)_n = \begin{cases}
 (c v_{n-3}+b v_{n-2} + a v_{n-1} + v_n )& n<0\\
\left[ (c v_{-3} +b v_{-2} + a v_{-1} +v_0)\right.& \\
  \quad \left. + (v_0 + a v_1 +b v_2+cv_3)\right] &n=0\\
 (  v_n + a v_{n+1} + b v_{n+2} +c v_{n+3}) & n>0
 \end{cases}
 \eeq
 has the solution $Mv=0$ with $v_n = x^{|n|}$, where $x$ is any of the three roots of the equation  $1+a x + b x^2 + c x^3 =0$.  The vector is normalizable if $|x|<1$, and three normalizable zero modes may be found over a range of parameters $a,b,c$.  The topological nature of the underlying theory is manifested by the fact that the the number of normalizable zero mode solutions does not change with small local excursions of $M$ away from the above form.
When such a system is gauged, however, the gauge fields couple to an infinite number of flavors in such a theory, and so the theory has a Landau pole and is ill-defined.  To cure this, one must work with a finite discretized extra dimension.

\begin{figure}[b]
\includegraphics[width=4.5 cm]{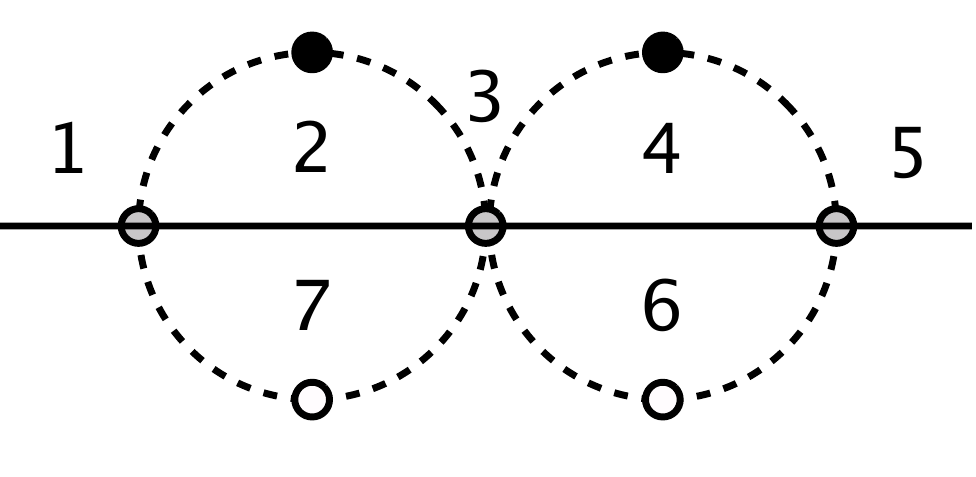}
\caption{\label{fig:sigmu} {A discretized $Z_2$ orbifold with three zero modes; $\CR$ reflects about the horizontal axis.}}
\label{fig:Z2orb}
\end{figure}

With finite continuous extra dimensions, both exponentially growing and falling zero mode solutions are normalizable and to obtain a chiral theory one must perform an orbifold projection \cite{Dixon:1985jw,Pomarol:1998sd}.  A well-known example is to compactify the extra dimension as a circle parametrized by $\theta\in[-\pi,\pi)$ with the action respecting a $Z_2$ symmetry $\psi(\theta)\to \gamma_5\psi(-\theta)$.  This symmetry requires mass terms to be odd in $\theta$, and so domain wall defects exist at the fixed points of the reflection, $\theta=0$ and $\theta=\pi$ where the zero modes of opposite chirality are located, and projecting out modes which are either even or odd under this $Z_2$ will result in a chiral theory.  A true UV completion of our topological insulator model for families is possible if a discretized version of this orbifold projection exists and is compatible with the survival of multiple chiral zero modes. In this case $M$ is a finite dimensional matrix, and we assume the action respects a $Z_2$ symmetry $\psi \to \gamma_5\CR\psi$, where $\CR$ is the ``reflection" operator in flavor space with $\CR^2=1$.  It is possible to prove an index theorem
\beq
(\CN_L^- - \CN_R^-) = -(\CN_L^+ - \CN_R^+)= \Tr\CR\ ,
\eeq
 where $\CN_L^\pm$ are the number of left-handed zero modes of $M$ with $Z_2$ charge $\pm1$, and $\CN^\pm_R$ is the same for right-handed zero modes.  Evidently we would like $|\Tr\CR|=3$ to obtain three chiral families, ruling out discretization of the $Z_2$ orbifold of a circle, and leading us to consider instead the discretization of an extra dimension consisting of two circles sharing a point.  A simple example is the 7-site lattice shown in Fig.~\ref{fig:Z2orb}, where we take 
$\CL_m$ to be
\beq
 &&\bar L_1 (R_2-R_7) + 
 \bar L_2(R_3-R_1) + 
 \bar L_3 (R_4-R_2) \cr &&+ \bar L_3( R_7- R_6) + \bar L_4( R_5- R_3) + 
 \bar L_6 ( R_3- R_5)\cr && + 
 \bar L_7 (R_1 -R_3)  + \text{h.c.}
\eeq
where $L_i$ and $R_i$ are $\psi_L$ and $\psi_R$ respectively at site $i$.  This has the simple interpretation of hopping terms for fermions around the circles, with fermions at the shared point (site 3) being able to hop onto either circle.
In this example, $\CR$  reflects sites about the horizontal axis in Fig.~\ref{fig:Z2orb}, and the action is invariant under $\psi\to \gamma_5\CR\psi$. The theory has three massless Dirac fermions, and if we project out all states with negative $Z_2$ charge, then we are left with three left-handed zero modes.  The important point of this model is not that $M$ has three zero mode solutions, but that the zero modes persist even if $M$ is perturbed in any random way which respects the $Z_2$ symmetry --- although the eigenvectors (``transverse wave functions") will be altered under this perturbation--- exactly as one would expect from the topological origin of this model.

 This mechanism in general --- and the deconstructed model in particular --- naturally suggests a number of interesting directions to pursue, most interestingly whether it imposes inescapable constraints on Higgs and CP physics that might be probed by the LHC, as well as observable flavor changing neutral currents or lepton flavor violation, which would be characterized by the scale of the extra dimension \footnote{S. Sun, in preparation.}.

  \begin{acknowledgments}

We thank W. Grimus, A. Karch, R. Sundrum and E. Witten for helpful comments. This work was supported in part by U.S.\ DOE grant No.\ DE-FG02-00ER41132.
\end{acknowledgments}
\bibliography{flavorbib}
\end{document}